\documentclass[prl,preprint]{revtex4}%
\usepackage{amsfonts}
\usepackage{amsmath}
\usepackage{amssymb}
\usepackage{graphicx}%
\providecommand{\U}[1]{\protect\rule{.1in}{.1in}}

\begin{document}
\title[Imbalanced Fermi superfluid in an optical potential]{Imbalanced Fermi superfluid in a one-dimensional optical potential}
\author{J. Tempere$^{\ast}$, M. Wouters and J.T. Devreese}
\affiliation{TFVS, Universiteit Antwerpen, Groenenborgerlaan 171, B2020 Antwerpen, Belgium}

\begin{abstract}
The superfluid properties of a two-state Fermi mixture in an optical lattice
are profoundly modified when an imbalance in the population of the two states
is present.We present analytical solutions for the free energy, and for the
gap and number equations in the saddle-point approximation describing resonant
superfluidity in the quasi-two-dimensional gas. Inhomogeneities due to the
trapping potentials can be taken into account using the local density
approximation. Analyzing the free energy in this approximation, we find that
phase separation occurs in the layers. The phase diagram of the superfluid and
normal phases is derived and analytical expressions for the phase lines are
presented. We complete the investigation by accounting for effects beyond
mean-field in the BEC limit where the system is more properly described as a
Bose-Fermi mixture of atoms and molecules.

\end{abstract}
\maketitle

The formation of pairing correlations in a mixture of two types of fermions is
frustrated when the number of fermions in each state is unequal. The effect of
such "spin imbalance" on pairing has been investigated theoretically since
Clogston's seminal work \cite{Clogston} for conventional superconductors, and
has led to the prediction of novel pairing states \cite{FFLO}, i.a. relevant
for color superconductivity in dense quark matter \cite{quark} and for
neutron-proton pairing in asymmetric nuclear matter \cite{nuclear}. The
experimental study of Fermi superfluids with imbalanced spin population has
only recently become possible, in ultracold atomic gases
\cite{MIT-imbal,RICE-imbal}, renewing theoretical interest in imbalanced
superfluidity \cite{theory-imbal,bedaque,IskinPRL97}.

In these experiments, the interaction strength between ultracold fermions can
be precisely controlled through the use of Feshbach resonances. These
scattering resonances allow to tune the s-wave scattering length from a large
negative value, giving rise to a Bardeen-Cooper-Schrieffer (BCS) superfluid
\cite{FermiBCS}, to a large positive value where a Bose-Einstein condensate
(BEC) of weakly bound molecules is formed \cite{FermiBEC}.

Also the trapping geometry confining the Fermi gas can be precisely controlled
experimentally. Of particular interest is the possibility to impose a
crystalline potential through the use of optical lattices. These allow to
experimentally mimic theoretical lattice models such as the Hubbard model.
One-dimensional optical lattices allow to study stacked layers of superfluid
and create a geometry analogous to layered (cuprate) high-temperature
superconductors. Optical lattices have been used to demonstrate superfluid
behavior of condensates \cite{CataliottiSCI293}, and to probe the
Mott-superfluid transition \cite{GreinerNAT415}. So far, fermionic
superfluidity in an optical lattice has only been studied \cite{MIT-optilatt}
with balanced Fermi gases.

We investigate the effect of spin imbalance on the superfluid properties of a
Fermi gas in an optical lattice. In particular, we will examine the case of a
one-dimensional optical lattice generated by two counterpropagating laser
beams (parallel to the $z$-axis) with wave length $\lambda$. These laser beams
generate a periodic potential $V_{0}\sin^{2}\left(  2\pi z/\lambda\right)
.$When loaded in this optical lattice, the gas forms a stack of typically a
few hundred quasi-2D layers containing several thousands of atoms each. The
interaction between atoms within a given quasi-2D layer can be modelled by a
2D contact interaction whose strength $g$ depends on the model cutoff $K_{c}$
and on the energy of the scattering atoms through \cite{RanderiaPRB41}%
\begin{equation}
\frac{1}{g}=\frac{m}{4\hbar^{2}}\left[  i-\frac{\ln\left(  E/E_{b}\right)
}{\pi}\right]  -\int_{k<K_{c}}\frac{d^{2}\mathbf{k}}{(2\pi)^{3}}\frac
{1}{(\hbar k)^{2}/m-E+i\varepsilon}.\label{g}%
\end{equation}
Here, $m$ is the mass of the atoms and $E_{b}$ is the energy of the bound
state that always exists in two dimensions, given by%
\begin{equation}
E_{b}=\frac{C\hbar\omega_{L}}{\pi}\exp\left(  \sqrt{2\pi}\frac{\ell_{L}}%
{a_{s}}\right)  ,\label{Eb}%
\end{equation}
with $a_{s}$ the (3D) s-wave scattering length of the fermionic atoms,
$\omega_{L}=\sqrt{8\pi^{2}V_{0}/\left(  m\lambda^{2}\right)  }$ and $\ell
_{L}=\sqrt{\hbar/(m\omega_{L})}$ and $C\approx0.915$ (cf. Ref.
\cite{PetrovPRA64}).

In this letter we derive an \emph{analytical }expression for the free energy
of the gas with density $n$ in a layer in the optical potential. Extremizing
this free energy with respect to the superfluid gap allows to set up and solve
the gap equation. We derive analytical expression for both the gap and the
chemical potential, for fixed imbalance. However, we will argue that the
imbalanced gas may be unstable with respect to phase separation into a
balanced superfluid and a halo of excess carriers, similar to the
three-dimensional case \cite{bedaque}, and we derive the phase diagram for the
gas, again retrieving analytic expressions for the phase boundaries.

\bigskip

The partition sum of the quasi-2D Fermi gas can be written as a path integral
over the exponential of the action functional $S$ for the fermionic fields
$\bar{\psi}_{k,\sigma},\psi_{k,\sigma}$ where $\sigma$ denotes the spin. We
write $k=\{\mathbf{k},\omega_{n}\}$ for the 2D wave number \textbf{k} and the
Matsubara frequency $\omega_{n}=(2n+1)\pi/\beta$ where $\beta=1/(k_{B}T)$ is
the inverse temperature:%
\begin{equation}
\mathcal{Z}=\int\mathcal{D}\psi\mathcal{D}\bar{\psi}\exp\{-\mathcal{S}\},
\end{equation}
with
\begin{equation}
\mathcal{S}=\frac{1}{\beta}\sum_{\omega_{n}}\int_{\mathcal{B}}\frac
{d\mathbf{k}}{(2\pi)^{2}}\text{ }\left\{
{\textstyle\sum\limits_{\sigma=\uparrow,\downarrow}}
\bar{\psi}_{k,\sigma}\left[  -i\omega_{n}+\mathbf{k}^{2}-\mu_{\sigma}\right]
\psi_{k,\sigma}+g\bar{\psi}_{k,\uparrow}\bar{\psi}_{-k,\downarrow}%
\psi_{-k,\downarrow}\psi_{k,\uparrow}\right\}  .\label{S1}%
\end{equation}
Here, we use units such that $\hbar=k_{F}=E_{F}=1$. The number of spin-up and
spin-down fermions is controlled through the chemical potentials
$\mu_{\uparrow}$ and $\mu_{\downarrow}$. The partition sum corresponding to
the action functional (\ref{S1}) can be calculated following the standard
procedure of introducing the Hubbard-Stratonovic decomposition, integrating
out the Grassman variables, and performing the saddle-point approximation
\cite{MichielPRA70,popov} with a constant gap $\Delta$. We therefore neglect
the possibility of two-dimensional FFLO type states with a modulated order
parameter \cite{FFLO,LOFF}. It is useful to express the results as a function
of $\mu=(\mu_{\uparrow}+\mu_{\downarrow})/2$ determining the total number of
fermions and $\zeta=(\mu_{\uparrow}-\mu_{\downarrow})/2$ expressing the
imbalance in chemical potentials. We find $\mathcal{Z}_{sp}=\exp\{-\beta
\Omega_{sp}\}$ with
\begin{equation}
\Omega_{sp}=-\frac{\Delta^{2}}{g}-\frac{1}{\beta}\int\frac{d^{2}\mathbf{k}%
}{(2\pi)^{2}}\left\{  \log\left[  2\cosh(\beta\zeta)+2\cosh(\beta
E_{k})\right]  -\beta\xi_{k}\right\}  .\label{freeE}%
\end{equation}
Here the $k$-integration is convergent at large wave vectors so that the wave
vector cutoff $K_{c}$ can be sent to infinity, $\Delta$ represents the pairing
gap, $\xi_{k}=k^{2}-\mu$ is the free particle spectrum, and $E_{k}=\sqrt
{\xi_{k}^{2}+\Delta^{2}}$ is the Bogoliubov spectrum. The saddle-point free
energy $\Omega_{sp}$ allows to derive the gap and number equations. A similar
result was derived for the homogeneous 3D gas by Iskin and Sa de Melo
\cite{IskinPRL97}. For the 2D case however, we succeeded to perform the
wave-number integrations analytically in the limit of temperature going to
zero:%
\begin{align}
8\pi\Omega_{sp} &  =-\Delta^{2}\ln\left(  E_{b}\right)  -\mu\left(  \sqrt
{\mu^{2}+\Delta^{2}}+\mu\right)  +\Delta^{2}\ln\left(  \sqrt{\mu^{2}%
+\Delta^{2}}-\mu\right)  -\dfrac{\Delta^{2}}{2}\nonumber\\
&  -2\zeta\left(  k_{b}^{2}-k_{a}^{2}\right)  +\xi_{b}\sqrt{\xi_{b}^{2}%
+\Delta^{2}}-\xi_{a}\sqrt{\xi_{a}^{2}+\Delta^{2}}+\Delta^{2}\ln\left(
\dfrac{\sqrt{\xi_{b}^{2}+\Delta^{2}}+\xi_{b}}{\sqrt{\xi_{a}^{2}+\Delta^{2}%
}+\xi_{a}}\right)  \label{fresult}%
\end{align}
where $\xi_{a,b}=k_{a,b}^{2}-\mu$ and for $\zeta>\sqrt{\Delta^{2}+(\min
[\mu,0])^{2}}$
\begin{equation}
\left\{
\begin{array}
[c]{l}%
k_{a}=\sqrt{\max[\mu-\sqrt{\zeta^{2}-\Delta^{2}},0]}\\
k_{b}=\sqrt{\mu+\sqrt{\zeta^{2}-\Delta^{2}}}%
\end{array}
\right.  \text{ }%
\end{equation}
For $\zeta\leqslant\sqrt{\Delta^{2}+(\min[\mu,0])^{2}}$ the terms containing
$k_{a}$ and $k_{b}$ in (\ref{fresult}) vanish. The Bogoliubov energy $E_{k}$
for quasiparticle excitations with wave vector $k=k_{a},k_{b}$ becomes equal
to the imbalance of the chemical potentials $\zeta$. All quasiparticle states
with $k\in\lbrack k_{a},k_{b}]$ have a Bogoliubov energy less than $\zeta$,
and contribute as normal state particles, rather than as superfluid
excitations, to the free energy.

Iskin and Sa de Melo \cite{IskinPRL97} solve the gap and number equations for
the imbalanced three-dimensional superfluid numerically with respect to
$\zeta,\mu$ and $\Delta$ for a given imbalance $\delta n/n$ and a given
interaction strength $g$. In the two-dimensional case, this procedure applied
to (\ref{freeE}), yields analytical results, summarized in Table \ref{tab1}.
However, the question can be raised as to whether the state with fixed $\delta
n/n$ is stable with respect to phase separation into a phase with a balanced
superfluid ($\delta n/n=0$) in the center of the trap and an imbalanced halo
of excess spin component surrounding it. Indeed, in the experiment of Ketterle
and co-workers \cite{MIT-shellimage} for a three-dimensional gas with a
nonuniform trapping potential, the excess spin component is expelled in a
shell surrounding a spin-balanced superfluid . Will this be similar for a 2D
gas within a layer of the optical potential ?

\begin{table}[tb]
$%
\begin{tabular}
[c]{l|cc}\hline\hline
& $E_{b}/2<\left(  \delta n/n\right)  ^{2}E_{F}$ & $\left(  \delta n/n\right)
^{2}E_{F}<E_{b}/2$\\\hline
$\Delta^{2}$ & $0$ & $2E_{b}E_{F}\left(  1-h\left(  E_{F}/E_{b}\right)
\dfrac{\delta n}{n}\right)  $\\
$\mu$ & $E_{F}$ & $E_{F}-\dfrac{E_{b}}{2}\left(  1-h\left(  E_{F}%
/E_{b}\right)  \dfrac{\delta n}{n}\right)  $\\\hline\hline
\end{tabular}
$\caption{Analytical solutions for ${\protect\small \Delta}^{2}$ and
${\protect\small \mu}$ at fixed imbalance ${\protect\small \delta n/n}$ and
binding energy ${\protect\small E}_{b},$ where $h\left(  x\right)
=\max\left[  \sqrt{2x},1\right]  $. For $E_{b}/(2E_{F})<(\delta n/n)^{2}$
superfluidity is suppressed. There is a qualitative change in the dependence
of $\Delta^{2}$ and $\mu$ on $E_{b}$ when $E_{b}$ becomes equal to $2E_{F}$;
this value can be intepreted as separating a 'weak coupling' from a 'strong
coupling' regime.}%
\label{tab1}%
\end{table}

If we consider a 2D trapping potential $V\left(  \mathbf{r}\right)  $ that
varies slowly on the length scales set by the interparticle distance
$\ell=n^{-1/2}$ and the superfluid gap $\ell_{\Delta}=\sqrt{\hbar^{2}/m\Delta
}$, we can apply the procedure outlined by De Silva and Mueller
\cite{DeSilvaPRA73} for the imbalanced gas in a regular three-dimensional
trap. That is, we calculate $n\left(  \mathbf{r}\right)  $ and $\delta
n\left(  \mathbf{r}\right)  $ in the local density approximation, where the
local average chemical potential equals $\mu=\mu_{0}-V\left(  \mathbf{r}%
\right)  $ and the difference between the chemical potentials $\zeta=\zeta
_{0}$ is constant in space for a spin-independent trapping potential:%
\begin{align}
n\left(  \mathbf{r}\right)   &  =-\left.  \partial\Omega_{sp}/\partial
\mu\right\vert _{\mu=\mu_{0}-V\left(  r\right)  ,\zeta=\zeta_{0}}\\
\delta n\left(  \mathbf{r}\right)   &  =-\left.  \partial\Omega_{sp}%
/\partial\zeta\right\vert _{\mu=\mu_{0}-V\left(  r\right)  ,\zeta=\zeta_{0}}.
\end{align}
The gap is then found by extremizing $\Omega_{sp}$ with respect to $\Delta,$
for $\mu$ and $\zeta$ fixed by the local density approximation, rather than
for $\delta n/n$ fixed. The validity of this procedure depends on whether the
local density approximation is justified; this need not be the case for all
experimental setups \cite{RICE-imbal}.

For $\zeta=0$, we find that, the free energy $\Omega_{sp}$ shows a single
minimum at $\Delta_{\text{bal}}=\sqrt{2E_{b}\left(  \mu+E_{b}/2\right)  }$
\cite{RanderiaPRB41}, where its value is $\Omega_{\text{bal}}=-\left(
\mu+E_{b}/2\right)  ^{2}/(4\pi)$. This minimum represents the superfluid state
(since $\Delta\neq0$). As the imbalance $\zeta$ is increased, $\Omega_{sp}$
develops a second minimum around $\Delta=0,$ representing the normal state.
The free energy of the normal state is $\Omega_{0}=-[(\mu+\zeta)^{2}\Theta
(\mu+\zeta>0)\left.  +(\mu-\zeta)^{2}\Theta(\mu-\zeta>0)\right]  /(8\pi).$ We
have verified numerically that no other minima of $\Omega_{sp}$ except the
ones at $\Delta=\Delta_{\text{bal}}$ and $\Delta=0$ occur.

Upon increasing $\zeta$ the free energy curves are only affected in the region
$\Delta<\zeta$. The minimum at $\Delta=\Delta_{\text{bal}}$, representing the
superfluid state, is therefore not affected by changes in $\zeta$ until it
ceases to exist: $\Omega_{\text{bal}}$ is independent of $\zeta$. From this it
follows that the superfluid state does not sustain imbalance ($\delta n=0$
since $\partial\Omega_{\text{bal}}/\partial\zeta=0$). The free energy of the
normal state, $\Omega_{0}$, does depend on $\zeta$, so that the normal state
supports imbalance, as expected. This scenario is similar to that for a 3D
imbalanced Fermi mixture \cite{bedaque}.

To evaluate the phase boundary between the balanced superfluid and the
imbalanced normal state, we should make a distinction between the cases
$\mu>\zeta$ and $\mu<\zeta$. For $\mu>\zeta$ the thermodynamic potentials of
the superfluid and normal state are equal for $\mu E_{b}=\zeta^{2}%
-(E_{b}/2)^{2},$ whereas for $\zeta>\mu>-\zeta$, they are equal for
$\mu=(\zeta-\sqrt{2}E_{b})/(\sqrt{2}-1)$. In the normal state, the chemical
potential of at least one of the spin components (up or down) should be
positive for there to be any particles at all. This means that $\mu+\zeta>0$
or $\mu-\zeta>0.$ For $\zeta>0,$ the first condition is easier to fulfill, so
that it determines the boundary in the phase diagram with the "empty phase"
(no particles).

\begin{figure}
[ptb]
\begin{center}
\includegraphics[
height=2.4992in,
width=3.314in
]%
{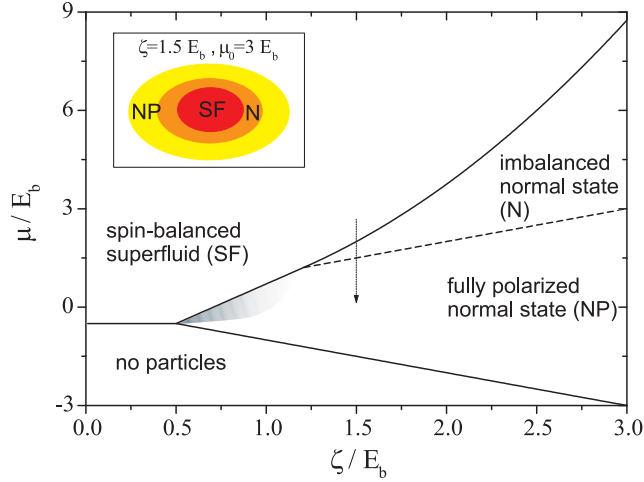}%
\caption{The phase diagram for the imbalanced (two-dimensional) Fermi gas is
shown as a function of the average chemical potential $\mu$ and the difference
between the chemical potentials $\zeta$. Three phases can be identified: a
balanced superfluid (SF), an imbalanced normal sate (N) and a fully polarized
normal state (NP). The arrow indicates the path in the phase diagram that is
traversed when moving away from the center of a 2D trap towards the edges. The
corresponding overall shell structure is illustrated in the inset. At very
strong coupling (large $E_{b}$) we expect corrections beyond mean field to
push the phase line down into the shaded region, where also a mixture of
bosonic molecules and fermionic atoms can appear.}%
\label{fig2}%
\end{center}
\end{figure}

Figure \ref{fig2} shows the phase diagram of the superfluid (SF), normal (N),
fully polarized normal (NP) and empty phases as a function of $\mu/E_{b}$ and
$\zeta/E_{b}.$ This phase diagram looks qualitatively similar to the one
obtained in 3D\ by mean field theory in Ref. \cite{DeSilvaPRA73}.%

Within this local density approximation, moving from the center of the trap
towards the edge corresponds to moving down along the arrow in figure
\ref{fig2}. From the center to the edge we encounter first the balanced
superfluid (SF), then a shell of spin-imbalanced normal gas, and finally a
shell of fully polarized normal gas, as illustrated in the inset. The
superfluid state in this treatment is balanced in the limit of temperature
zero, but allows for imbalance for any $T>0$.

In a typical experiment with a 1D optical lattice, several hundreds of layers
are present. As long as the Fermi energy and the superfluid gap are much
larger than the tunneling rate between the layers, the pairing is essentially
a phenomenon that takes place within each layer individually
\cite{giulianoPRL} and each layer can be treated in the local density
approximation by setting $\mu_{0}^{\left(  i\right)  }=\mu_{0}-U^{\left(
i\right)  },$ where $U^{\left(  i\right)  }$ is the trapping potential at the
center of the $i$th layer. The tunneling between the different layers is
between layers of zero imbalance, so that the results of Refs.
\cite{MichielPRA70,giulianoPRA71} can be used.

The preceding saddle point calculation implicitly makes the mean field
assumption that the typical size of the paired state $\ell_{\Delta}$ is much
larger than the distance between the fermionic atoms $\ell$. Deep in the BEC
limit, the system is in fact more suitably described as a Bose-Fermi mixture
of strongly bound bosonic pairs and excess majority component atoms. We again
can treat the 2D Bose-Fermi mixtures in the $\mu$-$\zeta$ plane analytically.
The energy density of the Bose-Fermi mixture is given by $E=2\pi n_{F}%
^{2}+g_{BB}n_{B}^{2}/2+g_{BF}n_{B}n_{F}$ where the density of bosonic
molecules $n_{B}$ and of fermionic atoms $n_{F}$ are positive. The coupling
constant $g_{BB}$ for dimer-dimer scattering in 2D has been calculated by
Petrov et al. \cite{petrov2D}, but not that for dimer-atom scattering,
$g_{BF}$. However in the limit of very low energy scattering, dimensional
arguments show that to first order the dimer-dimer and dimer-atom scattering
amplitudes are equal. The chemical potentials are $\mu_{F}=4\pi n_{F}%
+g_{BF}n_{B}$ for the fermionic atoms and $\mu_{B}=g_{BB}n_{B}+g_{BF}n_{F}$
for the bosonic molecules.

A minimal energy state has a positive curvature as a function of the
densities, i.e. the eigenvalues of $H_{ij}=\partial^{2}E/\partial
n_{i}\partial n_{j}\ $have to be positive. A necessary condition is that its
determinant is positive: $4\pi g_{BB}-g_{BF}^{2}>0.$ Inverting the expressions
for the chemical potential gives%
\begin{align}
n_{F}  &  =\frac{g_{BB}\mu_{F}-g_{BF}\mu_{B}}{4\pi g_{BB}-g_{BF}^{2}%
}\label{nf}\\
n_{B}  &  =\frac{4\pi\mu_{B}-g_{BF}\mu_{F}}{4\pi g_{BB}-g_{BF}^{2}} \label{nb}%
\end{align}
The expressions (\ref{nf}), (\ref{nb}) provide us with the information on the
phase diagram for phase separation of the 2D Bose-Fermi mixture. The condition
for a finite number of fermions is $\mu_{F}>g_{BF}\mu_{B}/g_{BB}$ and for a
finite number of bosons, it is $\mu_{B}>g_{BF}\mu_{F}/4\pi$, where we have
used $4\pi g_{BB}>g_{BF}^{2}$, as required for the stability of the system. To
use the above results for the BEC side of the BEC/BCS crossover with
imbalance, we substitute the bosonic chemical potential by $\mu_{B}%
\rightarrow\mu_{\uparrow}+\mu_{\downarrow}+E_{b}=2\mu+E_{b}$ and the fermionic
chemical potential by $\mu_{F}\rightarrow\mu_{\uparrow}=\mu+\zeta$. The
condition for a finite molecule density is then
\begin{equation}
\mu>-\frac{E_{b}}{2}\left(  1-\frac{g_{BF}}{2\pi}\right)  +\frac{g_{BF}}{4\pi
}\zeta\label{mmin}%
\end{equation}
while the condition for a finite fermion density reads%
\begin{equation}
\mu<\frac{\zeta-E_{b}g_{BF}/g_{BB}}{2g_{BF}/g_{BB}-1} \label{mmax}%
\end{equation}
In between these limits, a mixture of bosonic molecules and fermionic atoms
may coexist. Thus, we expect corrections beyond mean field to allow for an
additional phase to appear in the phase diagram of Fig.\ref{fig2}: the
imbalanced superfluid, qualitatively indicated by the shaded region. Moreover,
Eq. (\ref{mmax}) indicates that the line separating the spin-balanced
superfluid (molecules only) from the imbalanced superfluid lies below the mean
field SF-NP phase boundary of Fig.\ref{fig2}.

In conclusion, the experimental and theoretical investigation of resonance
superfluidity in Fermi gases has been marked by two recent advances: the
creation of a Fermi superfluid from a gas with unequal spin populations
\cite{MIT-imbal,RICE-imbal}, and the detection of fermionic superfluidity in
an optical lattice \cite{MIT-optilatt}. We have combined both effects and
studied the effect of unequal spin populations on the superfluidity in an
optical lattice.

We derived an analytical expression for the free energy of the imbalanced 2D
Fermi gas, eq. (\ref{fresult}). From this result, the BEC/BCS gap equation and
saddle-point number equations can be solved analytically, and results are
provided for fixed imbalance $\delta n$ in table (\ref{tab1}) However,
analyzing the free energy as a function of the superfluid gap we found that
phase separation occurs, and a (possibly imbalanced) normal and (always
balanced) superfluid phase separate. The phase boundaries are evaluated
analytically. The shell structure of the inhomogeneously trapped gas is
revealed to successively be from the center to the edges a balanced
superfluid, the spin-imbalanced normal state and the fully polarized normal
state. In the deep BEC limit a Bose-Fermi mixture offers a more suitable
description than the mean-field picture. In this Bose-Fermi mixture, an
imbalanced superfluid mixture of bosonic molecules and fermionic atoms is
possible in addition to the phases identified earlier.

\begin{acknowledgments}
Acknowledgments -- This research has been supported financially by the FWO-V
projects Nos. G.0356.06, G.0115.06, G.0435.03, G.0306.00, the W.O.G. project
WO.025.99N, the GOA BOF UA 2000 UA.. J.T. gratefully acknowledges support of
the Special Research Fund of the University of Antwerp, BOF\ NOI UA 2004 and
M.W. acknowledges his support from the FWO-Vlaanderen under the form of a
`mandaat postdoctoraal onderzoeker'.
\end{acknowledgments}

\end{document}